\documentclass[12pt]{iopart}
\usepackage{iopams,soul,color}
\usepackage{graphicx}

\newcommand{\unit}[1]{\,{\rm #1}}
\providecommand{\tabularnewline}{\\} %\usepackage{bm}% bold math

\begin{document}

\title[Unmagnetized dense plasma jet injection into a hot strongly magnetized
plasma]{Ideal
magnetohydrodynamic simulations of unmagnetized dense plasma jet injection 
into a hot strongly magnetized plasma}
\author{Wei Liu$^1$\footnote{Present address: Department of Radiation Physics,
The University of Texas M. D. Anderson Cancer Center, Houston, TX, USA, 77030}
and Scott C. Hsu$^2$}
\address{$^1$ Theoretical Division, Los Alamos National Laboratory, Los
  Alamos, NM, USA, 87545}
\address{$^2$ Physics Division, Los Alamos National Laboratory, Los Alamos, NM, USA,
87545}
\ead{scotthsu@lanl.gov}

\begin{abstract}
We present results from three-dimensional ideal magnetohydrodynamic
simulations of unmagnetized dense plasma jet injection into a uniform hot
strongly magnetized plasma, with the aim of providing insight into
core fueling of a tokamak with parameters relevant for ITER
%[M. Shimada et al., Nucl.\ Fusion~{\bf 47}, S1 (2007)] 
and NSTX (National Spherical Torus Experiment). 
%[M. Ono et al., Nucl.\ Fusion~{\bf 40}, 557 (2000)].
Unmagnetized dense
plasma jet injection is similar to compact toroid injection but with much higher
plasma density and total mass, and consequently lower required injection
velocity. Mass deposition of the jet into the background appears to be facilitated via
magnetic reconnection along the jet's trailing edge.
The penetration depth of the plasma jet into the background plasma is
mostly dependent on the jet's initial kinetic energy, and
a key requirement
for spatially localized mass deposition is for the jet's
slowing-down time to
be less than the time for the perturbed background magnetic flux to relax
due to magnetic reconnection.  This work suggests that more accurate treatment
of reconnection is needed to fully model this problem.
Parameters for unmagnetized dense plasma jet injection are identified
for localized core deposition as well as edge localized mode (ELM)
pacing applications in ITER and NSTX-relevant regimes.
\end{abstract}


\maketitle

\section{Introduction}
It is important to deliver fuel into the core of a tokamak fusion
plasma to maintain steady-state operation, achieve more efficient
utilization of the deuterium-tritium fuel, and optimize the energy
confinement time \cite{pp88}. The subject of plasma fueling and density
profile control is important for the successful operation of a future
reactor-grade tokamak such as ITER \cite{shimada07}. The injected fuel must have
sufficiently high directed energy to penetrate the strongly magnetized
edge plasma and reach the core \cite{vgp05,lhl09}.  A total particle
inventory of $10^{18}$--$10^{23}$ and a flow velocity of up to 800~km/s
(depending on injected density) are
required \cite{avg05}.  
It is desirable for plasmas to be deposited not only deeply but also
precisely in order to optimize bootstrap current and maintain optimized fusion
burn conditions \cite{avg05,rr08}.
Several fueling schemes have been proposed, such as gas injection,
pellet injection \cite{mfes79}, neutral beam injection, and compact
toroid (CT) injection \cite{phh88}.
All these methods will
have difficulties achieving localized core deposition in a reactor grade
tokamak plasma \cite{avg05}.

Injection of unmagnetized (or weakly magnetized) dense plasma jets is
an alternative to the above methods.
This is most similar to CT
injection in concept, but with the key advantages of having much higher
density (which allows lower injection velocities) and injector
hardware with a smaller footprint than CT
injector/accelerators (allowing for versatile placement of many injector units around
the tokamak). Unmagnetized dense plasma jets produced by a 
two-stage pulsed plasma source \cite{vh01} were 
successfully injected deeply into the spherical tokamak
Globus-M \cite{vgp05,avg05}.  In addition, very
recent developments in coaxial gun and mini-railgun technology
utilizing a pre-ionized fill plasma
\cite{wh10} (of any gas species) make it timely to use modern simulation tools to gain
insight into the penetration and injection dynamics of unmagnetized dense
plasma jets into an ITER or NSTX-like (National Spherical
Torus Experiment) \cite{ono00} plasma.
Unmagnetized dense plasma jet injection may also potentially find
applications for edge
localized mode (ELM) pacing \cite{lang07} and disruption mitigation \cite{bogatu08}.

Recent simulation results \cite{lhl09}
indicate that low $\beta$ CT injection has the potential to deposit fuel in a
precise manner at any point in the machine, from the edge to the core,
although a very high injection speed is required. Since a CT plasma is
confined by its magnetic field, its density is limited. The
plasma volume needs to be increased to accommodate a larger total
number of particles \cite{vgp05,avg05}.  
Preliminary simulation results \cite{lhl09} of unmagnetized dense plasma jets
injected into an ITER-relevant background plasma
suggest that deep but somewhat less-localized fueling (compared to CT's) is possible for
unmagnetized dense plasma jet injection. 
In this paper, we employ a simple
idealized model and three-dimensional (3D) ideal magnetohydrodynamic (MHD)
simulations of unmagnetized dense plasma jets
propagating into a uniform slab plasma
with uniform magnetic field perpendicular to the jet propagation
direction, mimicking jet fueling into an infinite aspect ratio
tokamak.  As we point out throughout this paper, an ideal MHD treatment
cannot capture all of the important dynamics of this problem accurately.  Our
aim in this work is to provide initial insight into the essential physics
occurring during unmagnetized dense plasma jet injection, and point the way toward
progressively more sophisticated physics models and simulation tools that will
be required, {\em e.g.},
resistive MHD and two-fluid models, as well as the inclusion of
realistic tokamak profiles and geometry (including
finite aspect ratio).  Most of the present results focus on
ITER-relevant parameters and injection velocities (Table~\ref{iter}).  In addition, some
preliminary results are discussed for NSTX-relevant parameters (Table~\ref{nstx}).

For completeness, we briefly place this work into the context
of a substantial body of research that has focused on
CT injection for tokamak refueling.  The conducting sphere (CS) theoretical model
was put forth some twenty years ago \cite{phh88,pp88,nw91} and described the injection
of a rigid conducting ball into a strongly magnetized background plasma.
In the CS model, the Alfv\'en speed of the background plasma is much higher
than the speed of the injected CT, and thus the background field can rearrange
itself virtually instantaneously in response to the injected CT\@.   In the CS
model, it is
the magnetic field pressure gradient of the background plasma that eventually
brings the injected CT to a stop.  More precisely, the injected CT stops when
the background magnetic energy excluded by the CT volume approximately equals
the CT's initial kinetic energy.   Later, Suzuki et al.~\cite{swsh00,shk00,shk01}
performed MHD numerical simulations taking into account the compressibility
of the injected plasma.  They found, by carefully analyzing the
energetics of the evolution of the injected CT, that the effects of plasma
compressibility are critical for slowing down the injected
plasma.  Plasma compressibility leads to the modification of both the magnetic
field and density at the leading and side interfaces between the
injected and background plasmas. Therefore the speed of localized
Alfv\'enic and acoustic dynamics of this interfacial region becomes of the same order
as the injection speed, and thus the perturbed background field and density no longer can
respond instantaneously to the injected plasma. This, in combination with the
higher injection velocities treated in our work, leads to the situation where
the background field can be ``stretched'' by the injected plasma, as shown both in
Suzuki et al.'s and our simulation results, with field line tension
thus playing a significant role in stopping the injected plasma. 
We note that our recent work \cite{lhl09} and the present work
extend the work of Suzuki et al.\ by treating higher ratios of background-to-injected
magnetic field strength ($>10$) and density (up to 1000), respectively, for the problem
of plasma injection into tokamaks.  Finally, within the context of
supersonic gas injection, the effects of jet heating (by the hot background tokamak
plasma) and polarization electric field (within the jet) on jet penetration
have been considered \cite{rozhansky06eps}.  Accurate modeling of
these effects is beyond the
capabilities of our compressible MHD code, and thus future work using a two-fluid code
with a better heating model is needed to definitively evaluate these effects.

The paper is organized as follows. Section~\ref{model} describes the
computational model and problem setup.  In Sec.~\ref{results},
simulation results on unmagnetized dense plasma jet injection and evolution
are presented for an ITER-relevant scenario (Table~\ref{iter}).
Conclusions and also implications for ELM pacing,
disruption mitigation, and NSTX-relevant (Table~\ref{nstx}) jet injection are
given in Sec.~\ref{conclusion}.

\section{Computational model}
\label{model}

\begin{figure}[!htp]
\begin{center}

  \scalebox{0.4}{\includegraphics{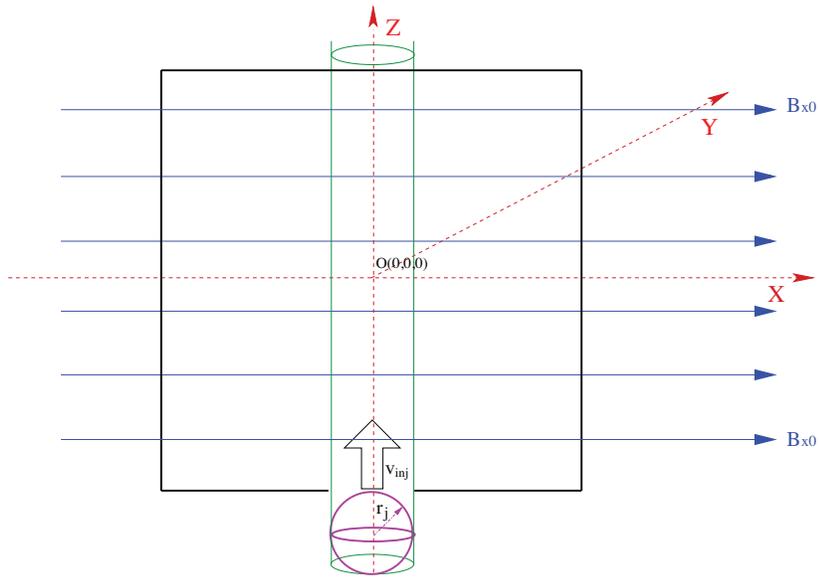}}
  \caption{\label{config}~(Color online) Schematic of the simulation
    geometry showing the coordinate system, the background magnetic field,
the injected plasma (with spherical radius $r_j=0.5$), and the background
plasma (cubic with sides of length $=18$).}
\end{center}
\end{figure}

An unmagnetized and high density ($\sim
10^{17}\unit{cm^{-3}}$) plasma jet with spherical radius $r_j=0.5$,
centered initially at $X=0$, $Y=0$ and $Z=Z_0=-10$, is
injected along the $Z$ axis into a lower density background plasma
with injection velocity $v_{\rm inj}$ (see Fig.~\ref{config}).  The background
plasma is a cube with sides of length $=18$.  We
use the term ``jet'' because eventually higher injected mass will
require that the ball become elongated into a cylindrical ``jet.'' The
model equations, assumptions, and numerical treatments are essentially the
same as those in a recent paper on CT injection \cite{lhl09},
except that the jet has spatially
uniform density rather than the double-peaked profile of a CT\@. 
The background plasma has a uniform magnetic field (no gradient).  However,
for the regimes considered in this paper, stopping of the injected plasma
is provided mostly by field line tension of the distorted background plasma,
and thus our results are expected to only slightly over-estimate the penetration
depth. Our 3D ideal MHD
code uses high-order Godunov-type finite-volume numerical
methods. These methods conservatively update the zone-averaged fluid
and magnetic field quantities based on estimated advective fluxes of
mass, momentum, energy, and magnetic field at the zone
interface \cite{llf06}. The divergence-free condition of the magnetic
field is ensured by a constrained transport scheme \cite{bs99}.
All simulations were performed on parallel Linux clusters at Los
Alamos National Laboratory.  We note that the details of 
magnetic reconnection and heat evolution are not captured
accurately due to the ideal MHD model and the use of a simplified
energy equation, and that future work using more sophisticated models are
needed to refine any reconnection-dependent conclusions given in this work.

Physical quantities are normalized by the characteristic system length
scale $R_0=10\;\unit{cm}$, mass density
$\rho_0=7.77\times10^{-9}\;\unit{g/cm^{-3}}$ (corresponding to a 50\%-50\% DT
mixture ion number density $n_0=1.86\times10^{15}\;\unit{cm^{-3}}$), and velocity
$V_{0}=1.7\times10^{8}\;\unit{cm\;s^{-1}}$. Other quantities are
normalized as (for the ITER-relevant case of Table~\ref{iter}):
time $t=1$ gives $R_0/V_0=59\unit{ns}$,
magnetic field $B=1$ gives
$(4\pi\rho_0V_{0}^2)^{1/2}=5.3\times10^{4}\;\unit{G}$, and energy
$E=1$ gives $\rho_0 V_{0}^2R_0^3=2.24\times10^{11}\;\unit{ergs}$. 
The boundary conditions are all perfectly conducting in the $Y$ and
$Z$ directions except at the entrance port of the bottom boundary
(analogous to the tokamak edge) where the jet is injected, while in
the $X$ direction non-reflecting
outflow boundary conditions \cite{rudy80} are employed in order to
mimic the toroidal geometry of a tokamak. 
The outflowing boundary
condition is stress-free, and thus the magnetic flux is not ``line-tied''
to the walls and Alfv\'en dynamics are
supported along $X$ despite the small simulation domain.
Suzuki et al.~\cite{shk01} have performed  simulations with toroidal
periodic boundary conditions.  They found that with such a stress-free
boundary condition, the magnetic tension force is smaller (but {\it
  not} zero), and the relaxation of the tension force from magnetic
reconnection is also much smaller (could even be ignored). An outflowing
boundary condition is more appropriate than Suzuki et al.'s choice for the large
aspect ratio case since in real tokamaks the toroidal dimension is
much larger than the poloidal dimension, and thus it will take a long
time for toroidal dynamics to re-enter the computational domain after
traveling along toroidal magnetic field lines \cite{lhl09}. In order to
minimize the influence of the entrance port on boundary conditions,
the port is opened (a hole inserted into the conducting boundary)
when the top of the jet reaches the bottom boundary at $t=0.5/v_{\rm
  inj}$ and is closed (hole removed from the conducting
boundary) at $t=2/v_{\rm inj}$ after the jet has fully entered the
computation domain (at $t\sim1.5/v_{\rm inj}$).  This leads to
an artificial force pulling back on the injected plasma which is 
apparent only at late times for simulations with shallow injection (as shown in
Fig.~7) and does not otherwise appear to 
significantly affect the injected plasma evolution.

In order to mimic jet injection into an ITER-relevant plasma, we adopt physical
quantities as given in Table~\ref{iter} in most of the results
reported in this paper. The major differences here compared to the recent results
on CT injection \cite{lhl09} are the very high density ratio 
$n_j/n_p\gtrsim500$ (where $n_j$ and $n_p$ are the injected and background
plasma densities, respectively) and the initially null value of the jet magnetic field.
The computational domain coinciding with the background plasma is $|X|\le9$,
$|Y|\le9$, and $|Z|\le9$, corresponding to a cube of $(180\;\unit{cm})^3$
in actual length units (assuming the
injected jet radius $r_j=5\;\unit{cm}$). The numerical resolution used
here is $400\times400\times800$, where the grid points are assigned
uniformly in the $X$, $Y$, and $Z$ directions. A cell $\delta X$
$(=\delta Y=2\delta Z=0.045$) corresponds to $0.45\;\unit{cm}$.  Since
the plasma skin depth and ion gyroradius based on either ITER or NSTX parameters
are no more than $\delta X$, simulations based on an MHD model are
reasonable for this initial study.

\section{Results}\label{results}

\subsection{Injected plasma jet evolution}
\label{evolution}

\begin{figure}[htbp]
\begin{center}
\scalebox{0.4}{\includegraphics{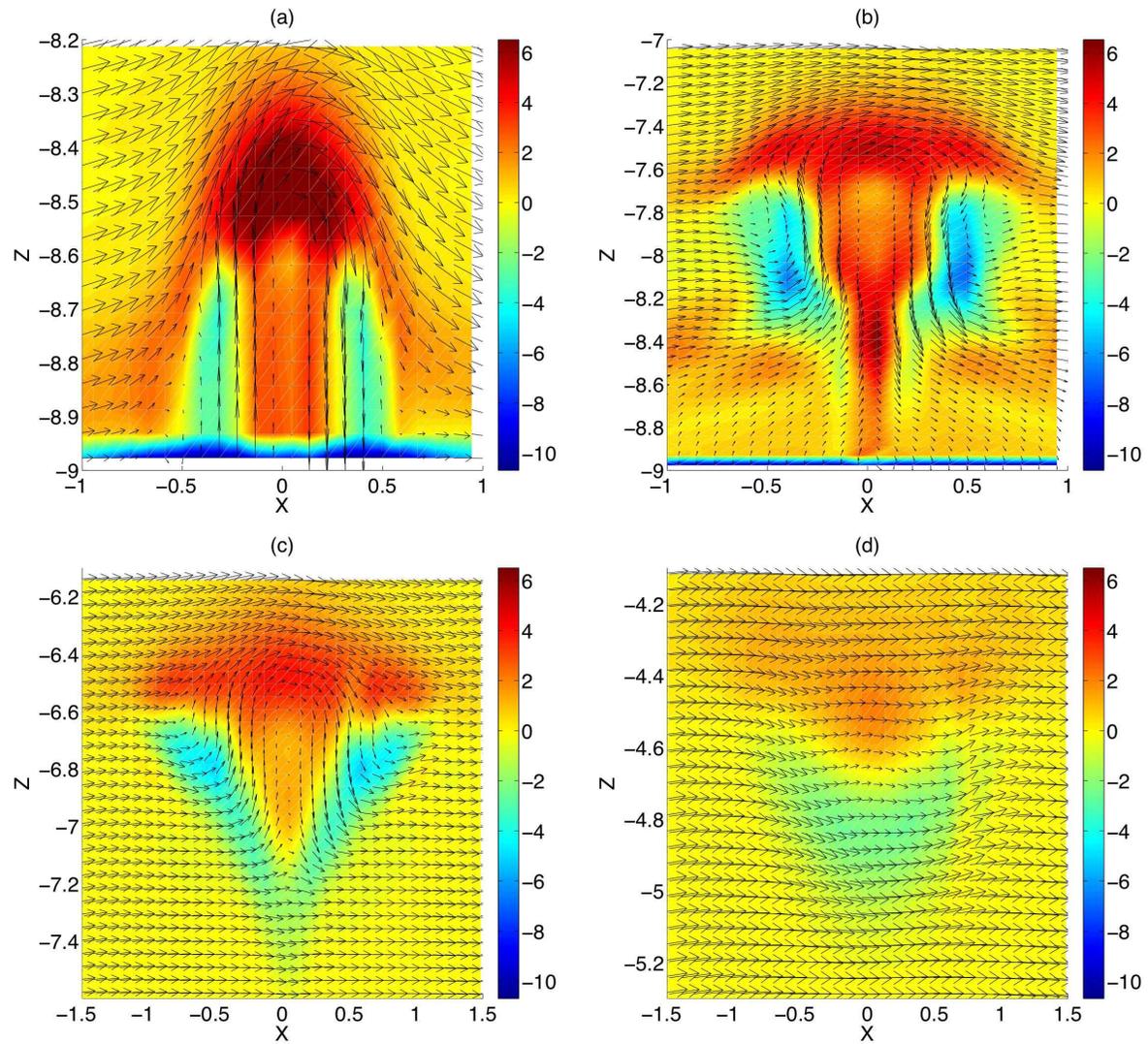}}
%\scalebox{0.25}{\includegraphics{j_vector40.eps}}$\;$
%\scalebox{0.25}{\includegraphics{j_vector80.eps}}$\;$\\ \scalebox{0.25}{\includegraphics{j_vector120.eps}}$\;$
%\scalebox{0.25}{\includegraphics{j_vector200.eps}}$\;$ 
\caption{(Color online) Contours of $j_y$ and vectors of
  $\vec{B}=(B_x,B_z)$ for parameters of Table~\ref{iter}:
(a)~$t=10$ ($0.59\unit{\mu s}$), (b)~$t=20$, (c)~$t=30$, (d)~$t=50$.
Note that the scales for the
  abscissa and ordinate are not identical.}
\label{reconnection}
\end{center}
\end{figure}

\begin{figure}[!htp]
\begin{center}

  \scalebox{0.6}{\includegraphics{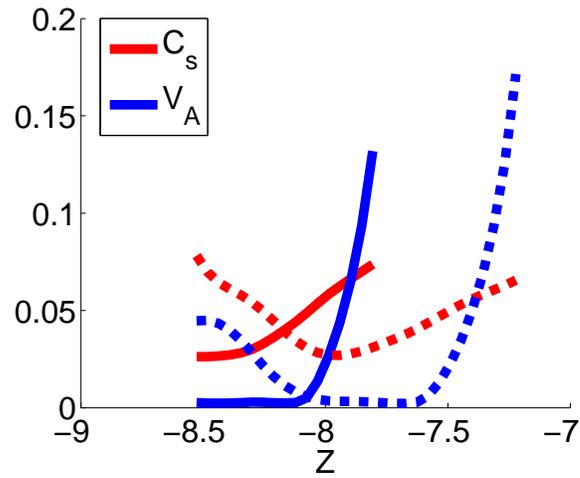}}
   \caption{\label{va-cs}~(Color online) Alfv\'en ($V_A$) and ion acoustic ($C_s$)
speeds calculated using local plasma parameters at $t=15$ (solid lines)
and $t=20$ (dashed lines), for the case of Table~1.}
\end{center}
\end{figure}

Figure~\ref{reconnection} displays the evolution of magnetic field $B_{xz}$
(arrows) and current density $j_y$ (color contours) in the $X$-$Z$
plane at $Y=0$ with parameters given in Table~1.
During the
initial jet penetration into the background plasma, the jet meets a very strong
magnetic barrier.  Thus, the plasma jet becomes compressed along the
direction of propagation ($Z$) by about 30\% at $t=20$ but not
much in $X$
(by examining the $\rho$ profile versus $Z$ and $X$ at $t=15$ and 20, not shown
here),
and the background magnetic field gets
distorted as seen in Fig.~\ref{reconnection}(a).
The plasma density increases at the jet-background interface.
A large plasma current also appears there
due to the compression of the background magnetic field, as
seen in Fig.~\ref{reconnection}(a) and the edges of the jet (return
current).  
Magnetic fields diffuse into the jet due to
numerical resistivity with resistive diffusion time
$\tau_{res}\sim 10.9$ \cite{lhl09}. For a real jet injected into
a real tokamak, the magnetic diffusion time into the jet
will likely be somewhat slower, and more sophisticated simulations
({\em e.g.}, using a resistive MHD or two-fluid code)
and ultimately experiments will be needed to fully assess this. 
However, we believe that the field diffusion into the jet is not
critically important for the further dynamics to be described below.
The magnetic tension force due to the stretched field lines 
persists even with our stress-free boundary conditions ({\em i.e.}, it is
not a result of line-tying), although it is smaller than
with fixed boundary conditions \cite{shk01}.
From Figs.~\ref{reconnection}(a) and (b),
the speed at which the background field distortion propagates away along
the $X$ direction can be crudely estimated as $\Delta X/\Delta t \sim
0.5/10 \sim0.05$, which is slower
than $v_{\rm inj}=0.12$.  In addition, Fig.~\ref{va-cs} shows the local values of
the Alfv\'en ($V_A$) and ion acoustic ($C_s$) speeds versus $Z$ at $t=15$ and $t=20$.
For example, at $t=20$, the jet/background interfacial region is centered
near $Z=-7.4$ as seen from Fig.~\ref{reconnection}(b), at which location
both $V_A$ and $C_s$ are of the order 0.05 as seen in Fig.~\ref{va-cs}.  Thus,
neither Alfv\'en nor acoustic waves are fast enough to instantaneously
relax the perturbations in magnetic field and density at the jet/background
interface.  Finally, tilting of the plasma jet is not observed 
due to the fast injection and short jet transit time.

\begin{figure}[!htp]
\begin{center}

  \scalebox{0.6}{\includegraphics{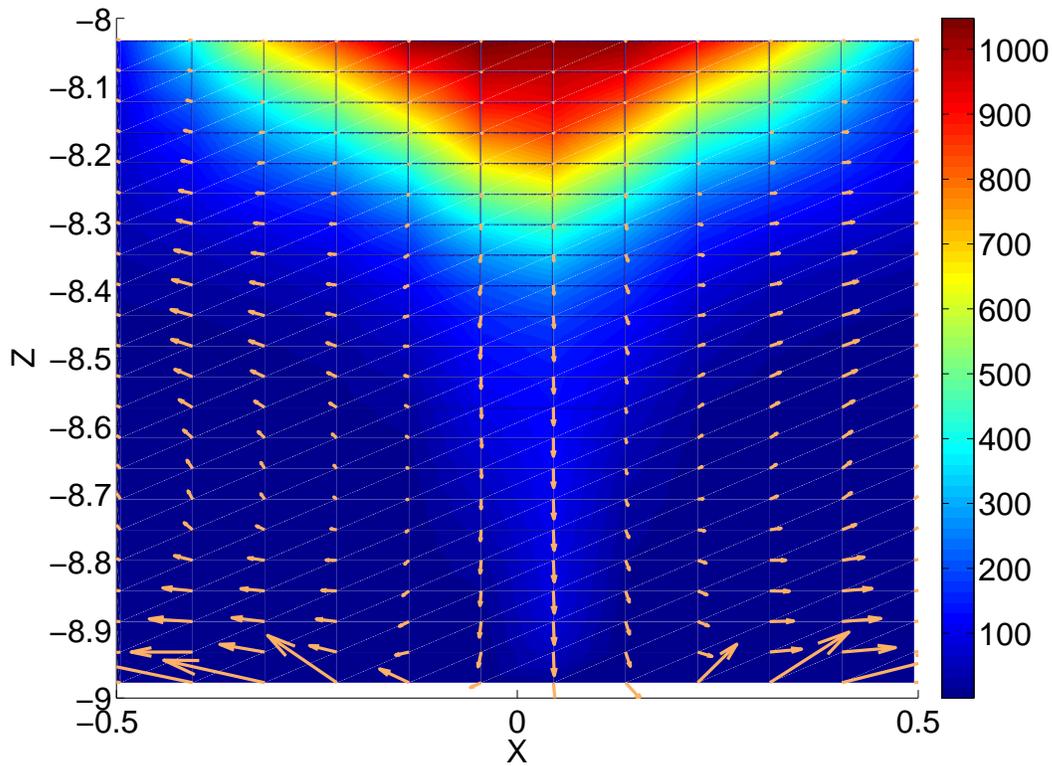}}
   \caption{\label{rho_vxz}~(Color online) Contours of $\rho$ and
vectors of $\vec{v}=(v_x,v_z)$ (in jet frame of reference)
at $t=20$ ($1.18\unit{\mu s}$) for parameters of Table~\ref{iter}.}
\end{center}
\end{figure}

\begin{figure}[!htp]
\begin{center}

  \scalebox{0.4}{\includegraphics{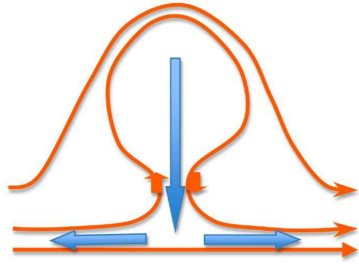}}
   \caption{\label{mechanism}~(Color online) Schematic of the 
     magnetic configuration around $t=20$.  Solid orange lines indicate magnetic
     field lines, while solid blue arrows indicate internal jet plasma flows
(in jet frame of reference).
     Opposing orange arrows indicate anti-parallel magnetic field
     where reconnection occurs.}
\end{center}
\end{figure}

After the jet has fully entered the background plasma region (after
$t\sim20$), a region with magnetic field reversal is set up
along the jet's trailing edge, centered about $X=0$ and $Z\approx -8.4$ in
Fig.~\ref{reconnection}(b). This enables magnetic reconnection \cite{yamada10}
which as pointed out by Suzuki et
al.~\cite{shk00} allows the jet to be ``detached'' from the background
field lines and by which the magnetic tension force
decelerating the jet is relaxed. Clearly, any reconnection observed in our
results is due to numerical resistivity (see further discussion in Sec.~\ref{depth}).
Figure~\ref{rho_vxz} shows velocity vectors,
from which [along with Fig.~\ref{reconnection}(b)] we infer a jet configuration at $t=20$
shown schematically in Fig.~\ref{mechanism}.  The primary
reconnection site is located at the trailing portion of
the jet along $Z$.  This
is qualitatively different from CT fueling in which reconnection takes place at the
upper left and lower right areas of the CT (due to the asymmetric CT
field) \cite{lhl09}. The reconnection process allows the plasma within
the jet to escape and eventually flow outward along the background
magnetic field horizontally (toroidally), as shown in Figs.~\ref{rho_vxz} and
\ref{mechanism}.   Thus reconnection facilitates
mass deposition from the jet into the background plasma.  
We note that mass deposition into the tokamak plasma
was indeed observed for unmagnetized plasma injection
on Globus-M \cite{vgp05,avg05} although neutralization of their jet
in transit could also lead to mass deposition without requiring reconnection.
The latter needs further study including atomic physics modeling.
The initial $Z$-directed kinetic energy is converted into $X$-directed kinetic
energy. 
After the high density jet plasma has been depleted (after $t\sim50$),
the perturbed background field has nearly
oriented again along $X$, the direction of the initial background magnetic field
[Fig.~\ref{reconnection}(d)]. 

\begin{figure}[!htp]
\begin{center}
%\scalebox{0.4}{\includegraphics{rho_contour600.eps}}$\;$
%\scalebox{0.4}{\includegraphics{rho_contour290.eps}}$\;$
\scalebox{0.4}{\includegraphics{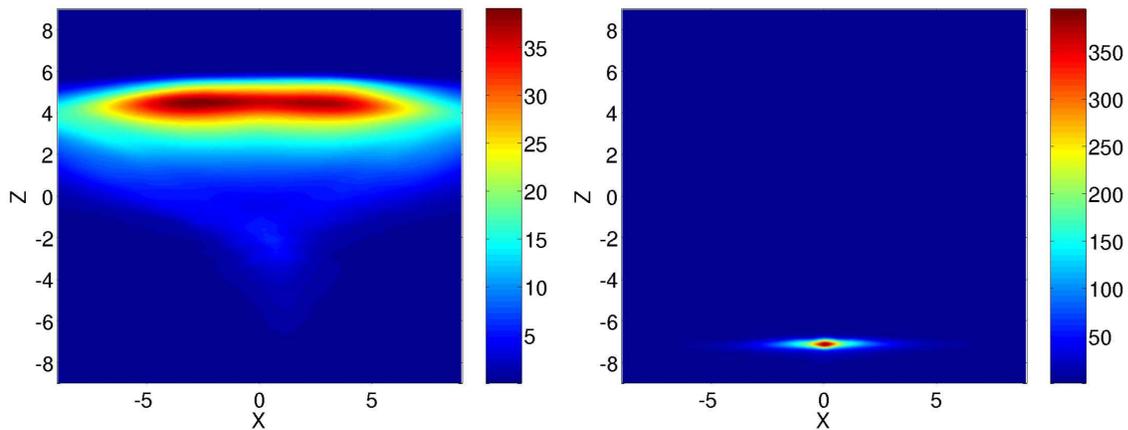}}
  \caption{\label{deep}~(Color online) (\emph{left}) Density contours
    at $Y=0$ and $t=150$ for parameters of Table~\ref{iter}, showing deep but
    not highly localized deposition.  The jet is still moving at nearly the
    initial injection speed. (\emph{right}) Density contours at $Y=0$
    and $t=80$ with the same parameters except $v_{\rm inj}=0.056$ ($93.3\unit{km/s}$),
showing shallow
    but more localized deposition. Note that the color scales are not identical.}
\end{center}
\end{figure}

Figure~\ref{deep}(\emph{left}) shows contours of density in the $X$-$Z$
plane ($Y=0$) at $t=150$ for the parameters of Table~1,
showing deep jet penetration. From Fig.~\ref{deep}(\emph{left}), the spread
$\Delta Z \sim3.5$ ($35\unit{cm}$) of this line-shaped structure is
much larger than for CT fueling \cite{lhl09} which is about
$\Delta Z \sim0.2$ ($2\unit{cm}$).  The jet in Fig.~\ref{deep}(\emph{left})
does not come to rest fully, but a jet with
lower injection velocity should come to
rest deep in the background plasma. 
Note that, by comparing with Fig.~2(d), the perturbed background magnetic flux
has mostly relaxed by $t=50$ while the diminishing
jet mass has reached $Z\approx 4$ as shown in Fig.~6(left).
However, under proper conditions (see further discussion in Sec.~\ref{depth}), the
jet mass can be
deposited more locally around the jet stopping position, as
seen in Fig.~\ref{deep}(\emph{right}) with lower
$v_{\rm inj}=0.056$.  A narrow elongated structure along $X$ with
penetration depth $2.7$ ($27\unit{cm}$) and a spread of $\Delta Z \sim0.5$
($5\unit{cm}$) results.
In order to
improve the precision of deposition, either lower injection speed,
lower initial jet mass, or larger
background magnetic field is necessary (see further discussion in Sec.~\ref{depth}).

\subsection{Penetration depth and localized deposition}\label{depth}

The penetration depth is mostly determined by the initial jet kinetic
energy. With the magnetic tension force from the background magnetic
field $F_{\rm tension}\propto B_p$, the deceleration of the jet
is $a_{\rm jet}=F_{\rm tension}/M_{\rm jet}$, where $M_{\rm jet}$ is
the jet initial mass. Note that Suzuki et al.~\cite{shk01} showed that
the CT penetration depth, based on a model with magnetic tension force
as the main deceleration mechanism, matches simulation results very
well, implying that MHD wave drag forces may not be important in CT
deceleration.  If the jet is undergoing a constant deceleration (a
rough estimate), the time for the jet to stop is: $T_{\rm
  slowdown}=v_{\rm inj}/a_{\rm jet}=v_{\rm inj}M_{\rm jet}/F_{\rm
  tension}$. Thus the penetration depth $S$ of the jet would be:
\begin{equation}
\label{penetration}
S=\frac{1}{2}v_{\rm inj}T_{\rm slowdown}=\frac{1}{2}M_{\rm jet}v_{\rm
  inj}^2/F_{\rm tension}=E_{K0}/F_{\rm tension}\;,
\end{equation}
where $E_{K0}=M_{\rm jet}v_{\rm inj}^2/2$ is the jet's initial
kinetic energy.  Based on this estimate, the penetration depth is
roughly proportional to the jet initial kinetic energy and inversely
proportional to the magnetic tension force, which is proportional to
the background magnetic field. The CS
model \cite{pp88} shows that a CT would penetrate to a position where
the initial CT kinetic energy exceeds the background magnetic field
energy excluded by the CT volume, and a lower limit of injection speed
$V_{\rm AC}$ was derived from this requirement \cite{lhl09}.
However, our simulation results [Fig.~\ref{deep}(\emph{left})]
with $v_{\rm inj}=0.12<V_{\rm
 AC}=0.136$ (for Table~1 parameters)
show that the jet penetrates the background field easily, which
was also observed experimentally \cite{avg05,vgp05}.
Instead, the jet must have
sufficiently high directed energy to overcome the deceleration from
the magnetic tension force. Of course, as mentioned previously,
another possibility is that a
sizable fraction of the initial jet plasma is transformed into
neutrals, which results in more efficient penetration of particles
into a magnetic field \cite{vgp05}.  A more detailed study including
the effects of atomic physics (especially three body recombination) is needed
to fully assess the role of jet neutralization during transit.

For internal density profile control, plasmas need to be deposited not
only deeply but also precisely \cite{avg05,rr08}. Based on the
mechanism of jet mass deposition put forth in Sec.~\ref{evolution}, a key requirement for
localized deposition is therefore to have the jet slowing-down time be less than
the time for the perturbed magnetic flux to relax due to magnetic
reconnection, $v_{\rm inj}M_{\rm jet}/F_{\rm tension}\lesssim\tau_{\rm
  res}$. Therefore lower injection speed (but exceeding the threshold
for penetration), smaller initial mass, and larger background magnetic
field are favorable for accurate deposition but unfavorable for
depositing high mass into the core of the tokamak.
Figure~\ref{deep}(\emph{right}) with $v_{\rm inj}=0.056$
($93\unit{km\;s^{-1}}$) and initial jet mass $0.22\unit{mg}$
shows a relatively shallow but local deposition case with 
penetration depth $S=2.7$ ($27\unit{cm}$) and spread $\Delta Z\sim0.5$
($5\unit{cm}$); this shallow deposition scenario may be useful for ELM pacing
applications (see
discussion in Sec.~\ref{conclusion}). Jet fueling might increase the
background plasma inventory by $\sim50\%$ in a single shot without
disturbing background plasmas parameters
[Fig.~\ref{deep}(\emph{right})].  Voronin et al.~\cite{vgp05} reported an
unmagnetized dense plasma jet source with an injection speed
$\sim110\unit{km\;s^{-1}}$. Coaxial guns and mini-railguns under
development by HyperV Technologies Corp.~\cite{wh10} have achieved peak
jet velocities of $100\unit{km\;s^{-1}}$, total masses up to
$4\unit{mg}$, and peak particle densities up to a few times
$10^{17}\unit{cm^{-3}}$, although all the
peak values have not yet been achieved simultaneously.  An unmagnetized dense
plasma jet with injection speed between $0.056$
($93\unit{km\;s^{-1}}$) and $0.12$ ($200\unit{km\;s^{-1}}$) with
particle number density $10^{17}$~cm$^{-3}$ is achievable
experimentally in the near term and might give deeper and localized
deposition.

Finally, due to the potentially
important role played by the reconnection time of the perturbed
magnetic flux for localized deposition,
we evaluate how the effective Lundquist number arising from numerical
resistivity in our ideal MHD simulation compares with the Lundquist number for the
situation in a real tokamak.
At $t=40$, the Alfv\'en speed $V_{A}$ and the thickness $L_{\rm
  sheet}$ of the current sheet at the leading edge of the jet are
found to be $7.5$ and $0.3$ from the simulation, respectively,
corresponding to an effective Lundquist number $=\tau_{\rm
  res}V_{A}/L_{\rm sheet}=250$ due to numerical resistivity. 
For the real situation, we assume Spitzer conductivity,
$\sigma=1.9\times10^4T_e^{3/2}Z\ln \Lambda\;\unit{(Ohm\;m)^{-1}}$,
where the electron temperature $T_e$ is in $\unit{eV}$ and the Coulomb
logarithm is
$\ln\Lambda=\ln(12\pi\epsilon_0^{3/2}T_e^{3/2}/Ze^{2/3}n_e^{1/2})$. The
magnetic diffusivity is $\eta=1/\mu_0\sigma$. Given
$n_e=1\times10^{17}\unit{cm^{-3}}$ and $T_e=2.5\unit{eV}$ as in
Table~\ref{iter}, we get $\eta=51\times10^{4}\unit{cm^2\;s^{-1}}$,
which gives a Lundquist number $=V_{A}L_{\rm sheet}/\eta=7500$ for the real situation.
Thus, the effective Lundquist number for jet-background interactions in our simulations
is much smaller than the likely real value in a
tokamak, meaning that our simulation results {\em underestimate the
reconnection time} of the perturbed magnetic flux and thus might {\em underestimate the
precision of deposition}.  The  latter statement needs to be verified by
more sophisticated
resistive MHD or two-fluid numerical modeling.  However, this will be challenging
because the many orders-of-magnitude difference in density and temperature in the problem
lead to huge differences in resistivity and viscosity.  An implicit MHD code
with small numerical viscosity {\em and} a Godunov scheme to handle shocks (these two
are conflicting requirements) is needed.

\section{Conclusion and Discussion}\label{conclusion}
In this paper results from 3D ideal MHD simulations
of unmagnetized dense plasma jet injection into a hot strongly magnetized
plasma are presented, with the aim of providing initial insight into core
fueling of a tokamak with parameters relevant for ITER\@. Unmagnetized
plasma jet injection is similar to CT injection but with higher
possible injection density and total mass, as well as a potentially
smaller footprint for the injector hardware.  Our simulation results
illustrate the jet evolution upon penetration of the background
plasma, suggesting that magnetic reconnection at the trailing
edge of the injected plasma jet plays an important role
in mass deposition.  Our results also show that
the penetration depth of the plasma jets is mostly dependent on
the jet's initial kinetic energy.
If the reconnection mechanism for mass deposition is correct, then a key
requirement for spatially localized fueling is for the jet slowing-down time to be
less than the time for the perturbed magnetic flux to relax due to
magnetic reconnection. Thus lower injection speed, smaller initial jet
mass, and larger background magnetic field favor precise
deposition. Proper conditions are identified for an unmagnetized dense
jet to have deep and localized deposition in an ITER-relevant plasma.
Future work including the use of 
resistive MHD and two-fluid models, realistic profiles
in both the background and injected plasmas, toroidal geometry of the background plasma,
and atomic physics effects ({\em e.g.}, three body recombination) potentially leading to
neutralization of the injected plasma jet during initial penetration, are needed
to refine the initial conclusions of this work based on ideal MHD.

\begin{figure}[!htp]
\begin{center}

  \scalebox{0.4}{\includegraphics{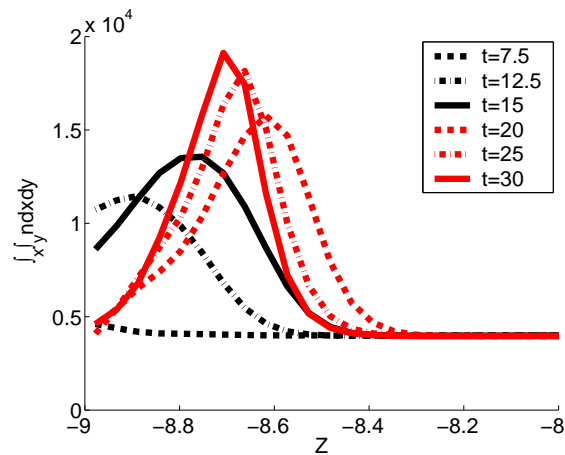}}
   \caption{\label{rhoz}~(Color online) Axial profiles of
     $\int_x\int_yn dxdy$ with $v_{\rm inj}=0.09$,
     $n_j/n_p=5.13$ (note different than Table~1),
 and $B_j/B_p=0$ for jet injected into ITER-like plasma
     (Table~\ref{iter}), showing shallow deposition of a large
     number of particles.  The ``bounce-back'' after $t=20$
is due to the open port boundary condition described in Sec.~2.}
\end{center}
\end{figure}

ELM mitigation via pacing is another potential application of
unmagnetized dense jet injection in ITER\@. 
To evaluate ELM pacing, we investigated the
case of an unmagnetized plasma jet with $n_j=5.13$
($9.5\times10^{15}\unit{cm^{-3}}$), $v_{\rm inj}=0.09$
($150\unit{km\;s^{-1}}$), and initial mass of $0.021\unit{mg}$
injected into a hot strongly magnetized background plasma with
ITER-relevant parameters (Table~\ref{iter})\@. Figure~\ref{rhoz}
displays the time evolution of the axial profile of $\int_x\int_yn
dxdy$.  The total number
of particles deposited near the bottom boundary (analogous to the
tokamak edge) is around $4.8\times10^{19}$, which is larger than the
total number of the particles needed ($4\times10^{19}$) for ELM
pacing \cite{lang07}.
The case in Fig.~\ref{rhoz} is
for a slow injection speed, but we have observed similar behavior for
faster injections (not shown here).
However, even by using argon with $n_j=53.8$
($1\times10^{17}\unit{cm^{-3}}$), the total mass would be only
$\sim3.5\unit{mg}$ mass per jet, which is about $\sim700$
times smaller than the needed mass $(\sim2.5\unit{g})$ for disruption
mitigation \cite{bogatu08}. Thus, the near term use of unmagnetized dense
jets for refueling and ELM pacing appear to be more promising than for
disruption mitigation.  One interesting proposal \cite{bogatu08} is the
use of heavy $C_{60}$-fullerene (buckyball) molecules by which dense jet
injection could potentially become a solution for disruption
mitigation. 

\begin{figure}[htbp]
\begin{center}
  \scalebox{0.2}{\includegraphics{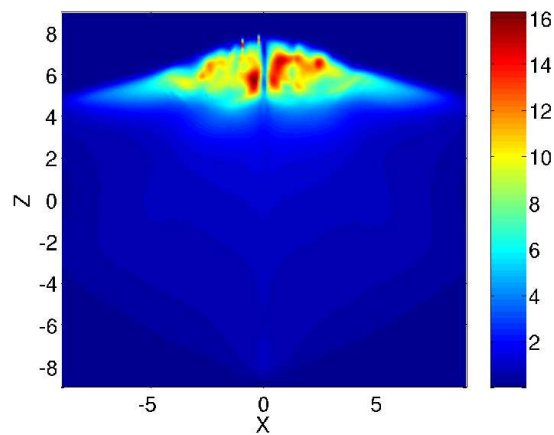}}$\;$
\caption{(Color online) \label{fig:nstx} Density  in the $X$-$Z$
    plane with $Y=0$ at $t=25$ ($8.53\unit{\mu s}$)
for NSTX-relevant parameters of Table~\ref{nstx},
showing deep jet penetration.  The density structure seen here is characteristic
of early time evolution likely due to internal jet mass flow induced by reconnection
at the tail.  The density smooths out at later times which is the case in Fig.~6.}
\label{intact}
\end{center}
\end{figure}

To explore and develop the unmagnetized dense plasma jet injection concept,
jets with parameters shown in Table~\ref{nstx} could be well
suited for fueling or ELM pacing applications on NSTX\@.  We have
performed a simulation of the injection of an unmagnetized
dense plasma jet with $n_j=100$ ($1\times10^{17}\unit{cm^{-3}}$),
$v_{\rm inj}=0.683$ ($200\unit{km\;s^{-1}}$) and jet initial
mass $0.17\unit{mg}$ (deuterium).  This simulation indicates
deep jet penetration but not very highly localized deposition (Fig.~\ref{fig:nstx}). 
Lower jet injection velocities which are presently
achievable could be
potentially used for ELM pacing which require only shallow penetration.
This preliminary result suggests
that NSTX would be a good platform to test the utility of unmagnetized
dense jet injection for the fueling and ELM pacing applications.
The required ITER fueling rate is $1.29\unit{mg}$ per shot at
$50\unit{Hz}$ \cite{om08}.   The parameters given in Table~\ref{iter}
are for a $0.22\unit{mg}$ jet.  Existing jets ({\em e.g.}, from HyperV
Technologies Corp.)~\cite{wh10}, as described in Sec.~\ref{depth},
already exceed these masses per jet.  The technology is simple enough
that a few of these guns could fire repetitively at up to
$10\unit{Hz}$ in concert to achieve the equivalent of $1.29\unit{mg}$
at $50\unit{Hz}$, although development will be needed for
a repetitive injection capability. Clearly, the mass and/or firing frequency
requirements for NSTX are much reduced.

\ack The authors thank Dr.~Shengtai Li for advice on the
code.  This work was funded by DOE
contract no.~DE-AC52-06NA25396 under the Los Alamos Laboratory
Directed Research and Development (LDRD) Program.

\section*{References}
%\bibliographystyle{unsrt}
%\bibliography{manuscript}

\newpage
\begin{table}[!htp]
\caption{\label{iter}~Initial parameters for the ITER-relevant
  case (DT plasma).  Jet injection velocity is $v_{\rm inj}=0.12$ ($200\unit{km/s}$).
The jet mass is $220\unit{\mu g}$.  One time unit corresponds to $59\unit{ns}$.}
\begin{tabular}{ccccc}

\hline  \hline & \multicolumn{2}{ c }{ Jet} & \multicolumn{2}{c }{
  Background} \tabularnewline \hline Parameter & numerical & physical
& numerical & physical \tabularnewline \hline Magnetic Field & 0
& 0 & $B_p=1.0$ &  $5.3\;\unit{T}$ \tabularnewline \hline Density
& $n_j=53.8$ & $1.0\times10^{17}\;\unit{cm^{-3}}$& $n_p=0.1$ &
$1.86\times10^{14}\;\unit{cm^{-3}}$ \tabularnewline \hline
Temperature & $T_j=3.33\times10^{-5}$ & $2.5\;\unit{eV}$ & $T_p=0.1$ &
$7.5\;\unit{keV}$ \tabularnewline  \hline  plasma $\beta=2n T/<B^2>$&
\multicolumn{2}{ c }{ $\infty$} & \multicolumn{2}{ c }{
  $\beta_p=0.02$}  \tabularnewline \hline \hline
\end{tabular}
\end{table}

\begin{table}[!htp]
\caption{\label{nstx}Initial parameters for the NSTX-relevant case (DD plasma).
Jet injection velocity $v_{\rm inj}=0.683$ ($200\unit{km/s}$).  The jet mass
is $170\unit{\mu g}$.  One time unit corresponds to $0.34\unit{\mu s}$.}
\begin{tabular}{ccccc}

\hline  \hline & \multicolumn{2}{ c }{ Jet} & \multicolumn{2}{c }{
  Background} \tabularnewline \hline Physical Quantities & numerical &
physical & numerical & physical \tabularnewline \hline Magnetic Field
& 0 & 0 & $B_p=1.0$ &  $0.6\;\unit{T}$ \tabularnewline %
\hline  Density & $n_j=100.0$ & $1.0\times10^{17}\;\unit{cm^{-3}}$&
$n_p=0.1$ & $1.0\times10^{14}\;\unit{cm^{-3}}$ \tabularnewline %
\hline Temperature & $T_j=1.67\times10^{-3}$ & $2.5\;\unit{eV}$ &
$T_p=1$ & $1.5\;\unit{keV}$ \tabularnewline  \hline  plasma $\beta=2n
T/<B^2>$& \multicolumn{2}{ c }{ $\infty$} & \multicolumn{2}{ c
}{ $\beta_p=0.2$}  \tabularnewline \hline \hline
\end{tabular}
\end{table}

\end{document}